\documentstyle[prl,aps,epsfig,prl,floats]{revtex}
\begin{document}

\draft 

\title{Dynamics and Thermodynamics of the Bose-Hubbard model}

\author{
  N.~Elstner and  H.~Monien
} 
\address{
  Physikalisches Institut, Universit\"at Bonn, 
  Nu\ss allee 12, D-53115 Bonn, Germany
}

\twocolumn[ \date{\today} \maketitle \widetext

\begin{abstract}
  \begin {center}
    \parbox{14cm}{ 
      We report results from a systematic analytic
      strong-coupling expansion of the Bose-Hubbard model in one and
      two spatial dimensions. We obtain numerically exact results for
      the dispersion of single particle and single hole excitations in
      the Mott insulator. The boundary of the Mott phase can be
      determined with previously unattainable accuracy in one and two
      dimensions. In one dimension we observe the occurrence of
      reentrant behavior from the compressible to the insulating phase
      in a region close to the critical point which was conjectured in
      earlier work. Our calculation can be used as a benchmark for the
      development of new numerical techniques for strongly correlated
      systems.}
  \end{center}
\end{abstract}

\pacs{\hspace{1.9cm} PACS numbers: 05.30.Jp, 05.70.Jk, 67.40.Db} ]

\narrowtext

Quantum phase transitions in strongly correlated systems have
attracted a lot of interest in recent years.  In fermionic systems the
Mott transition is complicated by the fact that in unfrustrated
systems the antiferromagnetic transition and localization transition
occur at the same point (see e.g. \cite{GeorgesReview:95}).  For
interacting Bose systems with spin zero the situation is much simpler
and one can focus on the physics of the Mott transition. Strongly
interacting bosonic systems are not only of academic interest.
Physical realizations include Josephson junction arrays, granular and
short-correlation-length superconductors, flux-lattices in type-II
superconductors and possibly in the future ultracold atoms in a
periodic potential \cite{Orr:86,Oudenaarden:96,Bruder:98}.

To be specific we investigate the generic model for the Mott
transition, the Hubbard model, for bosons (BH model):
\begin{equation}
H = -t\sum_{<i,j>} \left(b_i^\dagger b_j + b_j^\dagger b_i \right) 
    + \frac{1}{2} U \sum_i \hat n_i(\hat n_i - 1) - \mu \sum_i \hat n_i
\label{eq:BH-model}
\end{equation}
where the $b_i^\dagger$ and $b_i$ are bosonic creation and
annihilation operators, $\hat n_i = b_i^\dagger b_i$ is the number of
particles on site $i$, $t$ the hopping matrix element, $U>0$ the
on-site repulsion and $\mu$ the chemical potential. With short range
interactions only the model has two phases at zero temperature: a
superfluid phase and a Mott phase. Much of the physics of the model
was already understood qualitatively in an early paper by Fisher et
al.\cite{Fisher:89} and subsequent papers (see e.g.
\cite{Bruder:93,Roksar:91,Krauth:92}).

It is however interesting to obtain a quantitative understanding of
the model - for example to compare with experiments.  To this end the
BH model has been studied numerically by Quantum Monte Carlo
simulations
\cite{Scalettar:91,Batrouni:92,Batrouni:94,Niyaz:94,Krauth:91,vanOtterlo:94,Kashurnikov:96}
in one and two spatial dimensions. Recently the one-dimensional case
was also investigated using the density-matrix renormalisation group
(DMRG)\cite{Kuehner:97}.  This study found indications for an
unexpected reentrant behavior from the superfluid to the Mott-insulator as a
function of the hopping amplitude $t$ for certain values of the
chemical potential.

In this Letter we report for the first time a systematic analytic
strong coupling series to high order for the Bose-Hubbard model.
Previous attempts which were restricted to rather low order
\cite{Freericks:94-96} showed promising results but were not
sufficient to investigate the asymptotic behavior of the series.
Recently M. Gelfand\cite{Gelfand:96} proposed a method for a linked
cluster expansion with degenerate states.  We have implemented the
series expansion of the ground state and the first excited state as a
linked cluster expansion on a computer. The results show a spectacular
convergence of the Pade approximants for the phase diagram in one and
two dimensions.  The critical points can be determined to a previously
unattainable accuracy (relative errors of $\approx 10^{-3}$).  In
particular we are able to confirm convincingly that in one dimension
there is reentrant behavior of the Mott phase. The series calculation
can be used as a benchmark for development of new numerical techniques
for strongly correlated systems (e.g. DMRG).

We start by writing down the ground state in the atomic limit (the
hopping matrix element $t\rightarrow 0$). In the atomic limit
the number of bosons per site is fixed to an integer number say $n_0$.
Then the ground state of the
Mott insulator with a fixed number $n_0$ of particles per site is
given by
\begin{equation}
  \left| n_0 \right>_{\rm Mott}^{(0)} = \prod_{i=1}^N \frac{1}{\sqrt{n_0!}}
  \left(b^\dagger_i\right)^{n_0} \left|0\right> 
\end{equation}
with energy
\begin{equation}
  E_{\rm Mott}^{(0)}/N = \frac{1}{2} n_0(n_0 - 1) U - \mu n_0 \;\; .
\end{equation}
Perturbation theory for the ground state energy $E_{\rm Mott}$ can be
formulated as a linked cluster expansion, see e.g. \cite{Gelfand:90} and
the ground state energy can be obtained in the thermodynamic limit
``relatively easily''.

The Mott transition is obtained by studying charge excitations on top
of the Mott phase. The charge excitations are gapped in the
incompressible Mott phase and become gapless at the Mott transition.
In the atomic limit charge excitations are created by adding or
removing a particle onto or from a particular site $i$:
\begin{eqnarray}
  \left| n_0;i \right>_{\rm part}^{(0)} &=& 
  \frac{1}{\sqrt{(n_0+1)!}} \, b^\dagger_i 
  \left| n_0 \right>_{\rm Mott}^{(0)} \label{eq:sp-state} \\
  \left| n_0;i \right>_{\rm hole}^{(0)} &=& \frac{1}{\sqrt{n_0!}} \, b_i
  \left| n_0 \right>_{\rm Mott}^{(0)} 
  \label{eq:sh-state}
\end{eqnarray}
Their energy relative to the ground state is given by
\begin{equation}
  E_{\rm part/hole}^{(0)} = \pm \left( U n_0 - \mu \right) 
\end{equation}
for particles and holes respectively showing that the charge excitations are
degenerate. This degeneracy is lifted as soon as the hopping amplitude, $t$,
is finite. In the atomic limit the energy of the excited states vanishes for a
chemical potential $\mu_c^{(0)} = U n_0$ and the system becomes
compressible. .

A systematic strong coupling expansion of the energy of the charge
excitations complicated due to the high degeneracy. The problem how to
write down a linked cluster expansion for degenerate states was solved
only recently by Gelfand\cite{Gelfand:96}. The idea is to construct
perturbatively an effective Hamiltonian $H^{\rm eff}_{i,j}$ in the
subspace of the degenerate states $\left|n_0;i\right>_{\rm
  part/hole}^{(0)}$ by a similarity transformation
\begin{eqnarray}
H^{\rm eff}_{i,j}(t) &=& S_{i,\nu}(t) \, H_{\nu,\lambda} \, S_{\lambda,j}(t) \\
{\rm with} \phantom{W} S_{i,\nu}(t) &=& S_{\nu,i}^{-1}(t) \nonumber
\end{eqnarray}
where Greek indices run over states in the full Hilbert space while
Latin indices are restricted to the degenerate manifold of single
particle and single hole states (\ref{eq:sp-state}) and (\ref{eq:sh-state})
respectively. Then the linked cluster theorem applies to $H^{\rm
  eff}_{i,j}(t) - E_{\rm Mott}(t)$. In the case of a homogeneous system
$H^{\rm eff}_{i,j}$ depends only on the difference of indices $i-j$
and is easily diagonalised by a Fourier transform. This way one can
determine the full dispersion $E({\bf k}; t)$ of the charge
excitations. In many ways the linked cluster expansion is similar to a
exact diagonalization study of small systems - however in the linked
cluster expansion it is possible to remove all finite size effects in
each order and one obtains the full dispersion in the thermodynamic
limit.

For positive values of the hopping matrix element $t$ the smallest
(largest) eigenvalue in the particle (hole) sector is always located
at a wavevector ${\bf k} = 0$. The upper phase boundary of the Mott
phase is thus determined by the condition $E_{\rm hole}({\bf k=0};t) =
0$ and the lower boundary by $E_{\rm part}({\bf k=0};t) = 0$.  With
increasing hopping $t$ the distance between the upper and lower
boundary decreases until finally at some critical value, $t_c$, the
energy to remove a particle and the energy to add a particle become
degenerate and the Mott insulator vanishes altogether.

\begin{figure}[t]
  \begin{center} \epsfig{file=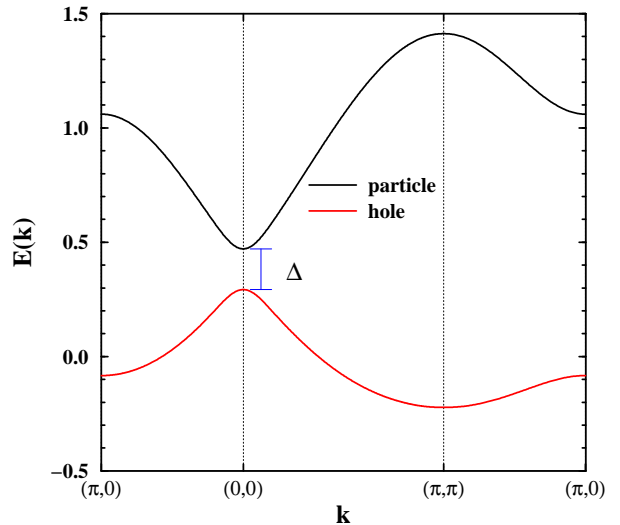,width=246pt} 
  \caption[The particle and hole excitation gap] 
          {Dispersion of the single-particle and single-hole excitations
           of the square lattice Bose-Hubbard model at $t/U=0.055$.}
  \label{fig:Dispersion} \end{center}
\end{figure}

We will first discuss the BH model, Eq. \ref{eq:BH-model} on a two
dimensional lattice. We investigated both the square and triangular
lattice and calculated the series for occupation numbers $n_0=1$ and
$n_0=2$ up to 12$^{\rm th}$ and 10$^{\rm th}$ order respectively.
The dispersion of the particle and hole excitations for $n_0=1$ on the
square lattice is shown in Fig.\ref{fig:Dispersion}. The different
shape of the two curves reflects the particle-hole asymmetry of the
model Hamiltonian (\ref{eq:BH-model}). The series were found to converge
very rapidly. Fig.\ref{fig:Dispersion} was obtained by summation of
the 12$^{\rm th}$ order series. It turned out to be almost
indistinguishable from the result of the 10-term series even for
$t/U=0.055$ which is not far from the critical endpoint $t_c$ of the
Mott lobe. The particle and hole excitations both have a pronounced
extremum at wavevector ${\bf k=0}$ and are separated by a gap
$\Delta$.  For values of the chemical potential $\mu$ in this range
all single charge excitations are gapped and the system is insulating.

\begin{figure}[t]
  \begin{center} 
    \epsfig{file=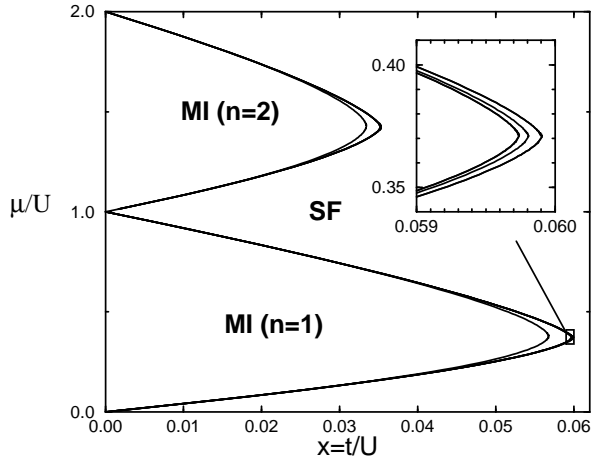,width=246pt} 
    \caption[Phase Diagram: Square Lattice] 
    { Phase diagram of the square lattice constructed using a Pade
      analysis of the series. The Mott phases are denoted by {\bf MI}
      and the superfluid region by {\bf SF}.
      The left curve is the lowest order Pade
      approximant ($4^{th}$ order series) the right curve represents all the
      higher approximants.  The inset shows a resolution of the region
      around the critical point. Note the scale!  }
    \label{fig:phase-diagram:2D}
  \end{center}
\end{figure}

The phase diagram shown in Fig. \ref{fig:phase-diagram:2D} is obtained
by a Pade analysis of the series for the single particle gap,
$\Delta$.  Scaling theory\cite{Fisher:89} predicts that in the
neighborhood of the critical point ($t_c$, $\mu_c$) the single
particle gap $\Delta(t)$ as a function of the hopping matrix element,
$t$, has the general form: $\Delta(t) = A(t) (t_c - t)^{z\nu}$ where
$A(t)$ is a regular function of $t$ and $z\nu$ is the dynamical
critical exponent. We use the following procedure to extrapolate the
series\cite{Guttmann}.  We calculate the logarithmic derivative of the
series of the gap with respect to $t$ which results in
\begin{equation}
  \label{eq:log-derivative}
  \frac{\partial\log(\Delta(t))}{\partial t} = 
  \frac{z\nu}{t - t_c} + \frac{A'(t)}{A(t)} 
\end{equation}
The right hand side of Eq. \ref{eq:log-derivative} is well
approximated by a Pade approximant. The pole of the Pade approximants
for ${\partial\log(\Delta(t))}/{\partial t}$ then determines the
critical point $t_c$ and the residuum determines the dynamical
critical exponent $z\nu$. We then integrate the Pade approximants
numerically to obtain the single particle gap $\Delta(t)$.  With the
exception of the lowest approximant all others approximant turn out to
be almost indistinguishable from each other indicating a rapid
convergence.  The results are shown in Fig.
\ref{fig:phase-diagram:2D}. To observe any change at all in the higher
approximants we have magnified the region around the critical point in
the inset. The chemical potential is a regular function of the hopping
matrix element $t$.  We used Pade analysis to check the scaling
prediction and found for the critical point $t_c \approx 0.0599 $ and
the critical exponent $\nu \approx 0.69$. This has to be compared with
the known value for the 3D xy-model\cite{Guillou:77}, $ \nu = 0.6693
\pm 0.0010 $, obtained by Borel summation of field theoretical
results.  The difference between the two results is of the order of a
few per cent. Obviously the Pade analysis has a tendency to slightly
overestimate the value of the critical point which in turn induces an
error in the value of the critical exponent.

\begin{figure}[t]
  \begin{center} 
    \epsfig{file=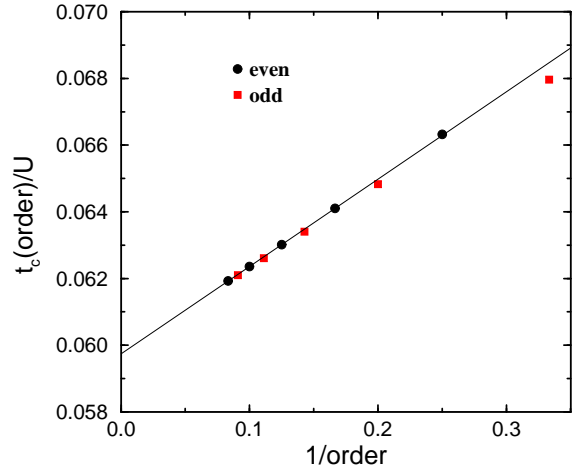,width=246pt} 
    \caption[The critical point] 
    {$1/order$ extrapolation of the critical point $(t_c)$.}
    \label{fig:crit-point} 
  \end{center}
\end{figure}

It is also possible to extract the critical hopping matrix element,
$t$, and the chemical potential at the critical point $\mu_c$ directly
from the series.  In each order $k$ of the expansion the single
particle gap $\Delta(t)$ vanishes at some effective critical value
$t_c^{(k)}$ with a corresponding effective $\mu_c^{(k)}$.  Plotting
$t_c^{(k)}$ and $\mu_c^{(k)}$ vs.  $1/k$ one finds again a rapid
convergence as shown in Fig.(\ref{fig:crit-point}). Extrapolation to
$k\rightarrow\infty$ allows to determine accurately the critical
point: $t_c = 0.05974 \pm 0.00004$ and $ \mu_c = 0.371 \pm 0.001$.

We now turn to the one-dimensional case. From scaling theory
\cite{Fisher:89} the critical behavior of the system is expected to be
that of a Kosterlitz-Thouless transition \cite{KosterlitzThouless} for
which the gap closes according to $ \Delta(t) \propto A(t) \exp\left(
  -{W}/{\sqrt{t_{\rm KT}-t}} \right)$ for $|t_{\rm KT}-t| \ll 1$,
where $A(t)$ is a regular function of $t$. The asymptotic form of the
gap makes it difficult to approximate $\Delta(t)$ directly. Therefore
we analyze the series for $\log(\Delta(t))^2$.
\begin{figure}[t]
  \begin{center} 
    \epsfig{file=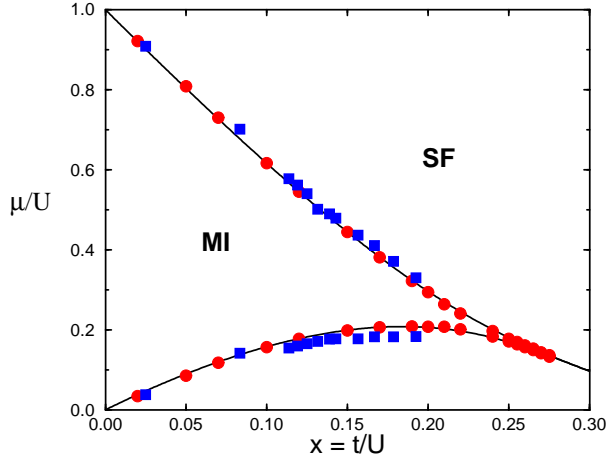,width=246pt} 
    \caption[Phase Diagram: 1D]  {
      Comparison of the phase diagram obtained from series expansion 
      (solid line), DMRG (solid circles) and QMC (solid squares).
      The Mott phase is denoted by {\bf MI} and the superfluid phase 
      by {\bf SF}.
      }
    \label{fig:phase-diagram:1D} 
  \end{center}
\end{figure}
The Pade analysis of the series yields spectacular agreement with the
recent DMRG study of the phase diagram \cite{Kuehner:97} as is shown
in Fig. \ref{fig:phase-diagram:1D} where we compare results from the
series analysis with numerical data of QMC simulations by Batrouni and
Scalettar\cite{Batrouni:94} and DMRG data\cite{Kuehner:97}. The
agreement between the series and the DMRG data is excellent. Both
calculations show that for a fixed chemical potential as a function of
the hopping matrix element $t$ the Mott phase is reentrant meaning
that by increasing the kinetic energy one returns to a {\bf localized}
state!  The series analysis confirms the surprising behavior observed
in the DMRG\cite{Kuehner:97} calculation. A simple intuitive way of
understanding this phenomenon is the fact that Mott lobe is
particle-hole asymmetric for the lattice problem. The
Kosterlitz-Thouless behavior at the phase transition then implies a
reentrant Mott phase. In higher dimensions this feature can not exist
because the transition has a power-law behavior. The accuracy of
previous calculations was not sufficient to observe the reentrant 
behavior.
The uncertainties in the precise location of the Kosterlitz-Thouless
transition are still comparatively large. We use a Pade analysis of
$\ln^2 \Delta(t) \propto (t_{\rm KT}-t)^{-1}$.  This quantity has a
simple pole at the critical point which can be captured by rational
function.  This methods turned out to give excellent results.  We
estimate the point for Kosterlitz-Thouless transition to be located at
$t_{\rm KT}/U = 0.26\pm0.01$ and $\mu_{\rm KT}/U = 0.16\pm0.01$.

In real systems disorder plays an important role. With our method it
is still possible to determine boundary of the Mott phase in that case
by asking where the system becomes compressible. For a detailed
discussion we refer the reader to Ref.  \cite{Freericks:94-96}.  The
``Bose-glass'' phase can not be studied with the techniques used above
- since the groundstate has no gap. 

In conclusion, series expansion techniques were applied to investigate
the zero temperature properties of the Bose-Hubbard model in one and
two dimensions. We determine the complete spectrum of single-particle
and single-hole excitations in the Mott phase. The phase diagram in
one and two dimensions is obtained {\em quantitatively} and the
critical end points of the Mott insulator regions are determined. In
two dimensions this is so far the only quantitative investigation of
the complete phase diagram of this problem.  In one dimensions the
series shows almost perfect agreement with a recent DMRG study and
provides a conclusive confirmation for counterintuitive reentrance
behavior from the compressible to the insulating phase near the
Kosterlitz-Thouless point. 

We acknowledge useful and interesting discussions on this problem with
M.~P.~Gelfand, T.~Giamarchi, A.~J.~Millis, A.~v.~Otterlo, 
R.~R.~P.~Singh, G.~Sch\"on, H.~Schulz.


\begin{references}
  \bibliographystyle{prsty}
  
\bibitem{GeorgesReview:95} A. Georges, G. Kotliar, W. Krauth and M. J.
  Rozenberg, Rev. Mod. Phys. {\bf 68}, 13 (1996)

\bibitem{Orr:86} B.~G. Orr, H.~M. Jaeger, A.~M. Goldman,
  and C.~G. Kuper, Phys. Rev. Lett. {\bf {\bf 56}}, 378 (1986); D.~B.
  Haviland, Y. Liu, and A.~M. Goldman, Phys. Rev. Lett. {\bf {\bf
      62}}, 2180 (1989); H.~M. Jaeger, D.~B. Haviland, B.~G. Orr, and
  A.~M. Goldman, Phys. Rev. B {\bf {\bf 40}}, 182 (1989).
      
\bibitem{Oudenaarden:96} A. Oudenaarden and J.~E. Mooij, Phys. Rev.
  Lett. {\bf {\bf 76}}, 4947 (1996).
  
\bibitem{Bruder:98}
  D. Jaksch, C. Bruder, J. I. Cirac, C. W. Gardiner and P. Zoller,
  cond-mat/9805329

\bibitem{Fisher:89} M.~P.~A. Fisher, P.~B. Weichman, G. Grinstein,
  and D.~S. Fisher, Phys. Rev. B {\bf {\bf 40}}, 546 (1989).

\bibitem{Bruder:93} C. Bruder, R. Fazio and G. Sch\"on, Phys. Rev. {\bf B47}, 
  342 (1993)

\bibitem{Roksar:91} D. S. Rhoksar and G. Kotliar,
  Phys. Rev. {\bf B44}, 10328 (1991)

\bibitem{Krauth:92} W. Krauth, M. Caffarel and J. P. Bouchard,
  Phys. Rev. {\bf B45}, 3137 (1992)

\bibitem{Scalettar:91} R.~T.~Scalettar, G.~G.~Batrouni and G.~T.~Zimany,
  Phys. Rev. Lett {\bf 66}, 3144 (1991)

\bibitem{Batrouni:92} G.~G.~Batrouni and R.~T.~Scalettar,
  Phys. Rev. B {\bf 46}, 9051 (1992)
  
\bibitem{Batrouni:94} G.~G.~Batrouni, R.~T.~Scalettar, G.~T.~Zimany
  and A.~P.~Kampf, Phys. Rev. Lett {\bf 72}, 3598 (1994)
     
\bibitem{Niyaz:94} P. Niyaz, R.~T. Scalettar, C.~Y. Fong and
  G.~G.~Batrouni, Phys. Rev. B {\bf 50}, 362 (1994)
  
\bibitem{Krauth:91} W. Krauth and N.~Trivedi, Europhys. Lett {\bf
    14}, 627 (1991), W. Krauth, N.~Trivedi and D.  Ceperley, Phys.
  Rev. Lett {\bf 67}, 2307 (1991)
  
\bibitem{vanOtterlo:94} A.~van~Otterlo and K.-H. Wagenblast, Phys. Rev. Lett
  {\bf 72}, 3598 (1994)
   
\bibitem{Kashurnikov:96} V.~A. Kashurnikov and B.~V. Svistunov,
  Phys. Rev. B {\bf {\bf 53}}, 11776 (1996).
    
\bibitem{Kuehner:97} T.~K\"uhner. diploma thesis, Bonn University,
  and T.~K\"uhner, and H.~Monien, preprint, cond-mat/9712307
  
\bibitem{Freericks:94-96} J.~K. Freericks and H. Monien, Europhys.
  Lett. {\bf \bf 26}, 545 (1994) and J.~K. Freericks and H. Monien,
  Phys. Rev. {\bf B53}, 2691 (1996).

\bibitem{Gelfand:96} M.~P.~Gelfand, Sol.~Stat.~Com. {\bf 98}, 11 (1996)
                 
\bibitem{Gelfand:90} M.~P.~Gelfand, R.~R.~P.~Singh, and D.~A.~Huse,
  J.~Stat.~Phys. {\bf 59}, 1093 (1990)
  
\bibitem{Guttmann}
  A.J. Guttmann,
  in 'Phase Transitions and Critical Phenomena' Vol. 13,
  C. Domb and J.L. Lebowitz editors

\bibitem{Guillou:77} J.~C.~Le ~Guillou, and J.~Zinn-Justin, Phys.~Rev.~Lett.
  {\bf 39}, 95 (1977)

\bibitem{KosterlitzThouless} V.~L. Berezinskii, Zh. Eksp. Teor. Fiz.
  {\bf {\bf 61}}, 1144 (1971) [JETP {\bf {\bf 34}}, 610 (1972)]; J.~M.
  Kosterlitz and D.~J. Thouless, Journal of Physics C {\bf {\bf 6}},
  1181 (1973); J.~M. Kosterlitz, Journal of Physics C {\bf {\bf 7}},
  1046 (1974).
  

\end{references}
\end{document}